\documentclass[showpacs,twocolumn,aps,preprintnumbers,letterpaper]{revtex4}
\usepackage{amsmath,amssymb}
\usepackage{epsfig}
\usepackage{graphicx}
\usepackage{amsmath}
\usepackage{slashed}
\usepackage{amsfonts}
\usepackage{epstopdf}

\newcommand{\be}{\begin{equation}}
\newcommand{\ee}{\end{equation}}
\newcommand{\bear}{\begin{eqnarray}}
\newcommand{\eear}{\end{eqnarray}}
\newcommand{\ba}{\begin{array}}
\newcommand{\ea}{\end{array}}

\def\be{\begin{eqnarray}}
\def\ee{\end{eqnarray}}
\def\bea{\be}
\def\eea{\ee}

\newcommand{\e}{{\mbox{e}}}

\def\roughly#1{\mathrel{\raise.3ex\hbox{$#1$\kern-.75em%
\lower1ex\hbox{$\sim$}}}}

\begin{document}

\title{The  Instanton-Dyon Liquid Model III:\\
 Finite Chemical Potential}

\author{Yizhuang Liu}
\email{yizhuang.liu@stonybrook.edu}
\affiliation{Department of Physics and Astronomy, Stony Brook University, Stony Brook, New York 11794-3800, USA}

\author{Edward Shuryak}
\email{edward.shuryak@stonybrook.edu}
\affiliation{Department of Physics and Astronomy, Stony Brook University, Stony Brook, New York 11794-3800, USA}

\author{Ismail Zahed}
\email{ismail.zahed@stonybrook.edu}
\affiliation{Department of Physics and Astronomy, Stony Brook University, Stony Brook, New York 11794-3800, USA}


\date{\today}
\begin{abstract}
We discuss an extension of the instanton-dyon liquid model that includes  light quarks at finite chemical potential
in the center symmetric phase. We develop the model in details for the case of $SU_c(2)\times SU_f(2)$ by mapping the 
theory on a 3-dimensional quantum effective theory. We analyze the different phases in the mean-field approximation.
We extend this analysis to the general case of $SU_c(N_c)\times SU_f(N_f)$ and note that the chiral and diquark pairings
are always comparable.
\end{abstract}


\maketitle

\setcounter{footnote}{0}


\section{Introduction}

This work is a continuation of our earlier studies~\cite{LIU1,LIU2}  of the  gauge 
topology in the confining phase of a theory with the simplest gauge group $SU(2)$.  
We suggested that the confining phase below the transition temperature is an
   ``instanton dyon" (and anti-dyon) plasma which is dense enough
to generate strong screening. The dense plasma is amenable to 
standard mean field methods.  
  
The treatment of the gauge topology near and below $T_c$ is based on the discovery of KvBLL
instantons threaded by finite holonomies~\cite{KVLL} and their splitting into 
the so called instanton-dyons (anti-dyons), also known as instanton-monopoles or instanton-quarks.
 Diakonov and Petrov and others \cite{DP,DPX}  suggested that the back reaction of the dyons on the holonomy potential 
at low temperature may be at the origin of the order-disorder transition of the Polyakov line.
Their model was based on (parts of) the one-loop
determinant providing the metric of the moduli spaces in BPS-protected sectors, purely selfdual or antiselfdual.
The dyon-antidyon interaction is not BPS protected and appears at the leading -- classical -- level,
related with the so called streamline configurations~\cite{LARSEN}.

The dissociation of instantons into constituents was advocated by Zhitnitsky and others~\cite{ARIEL}.
Using controlled semi-classical techniques on $S^1\times R^3$, Unsal and his collaborators~\cite{UNSAL} have shown that the repulsive interactions between pairs of dyon-anti-dyon (bions) drive the holonomy effective potential
to its symmetric (confining) value.

Since the instanton-dyons carry topological charge, they should have zero modes as well.   On the other hand, 
  for an arbitrary number of colors $N_c$ those topological charges are fractional $1/N_c$, while the
 number of zero modes must be integers. Therefore
  only some instanton-dyons may have zero modes~\cite{KRAAN}. For general $N_c$
 and general periodicity angle of the fermions the answer is known but a bit involved. For 
 $SU(2)$ colors and physically anti-periodic fermions the twisted $L$ dyons have zero modes,
 while the usual $M$-dyons do not.  Preliminary studies of the dyon-anti-dyon vacuum in the presence
 of light quarks were developed in~\cite{SHURYAK1,SHURYAK2}. In supersymmetric QCD some
 arguments were presented in~\cite{TIN}.

In this work we would like to follow up on our recent studies in~\cite{LIU1,LIU2} by switching a finite
chemical potential in the center symmetric phase of the instanton-dyon ensemble with light
quarks. We will make use of a mean-field analysis  to describe the interplay of the spontaneous
breaking of chiral symmetry with color superconductivity through diquark pairing. 
One of the chief achievement of this work is
to show how the induced chiral effective Lagrangian knows about confinement
at finite $\mu$. In particular, we detail the
interplay between the spontaneous breaking of chiral symmetry, 
the pairing of diquarks and center symmetry.

Many model studies of QCD at finite density have shown a competition between pairing of
quarks~\cite{BCS}, chiral density waves~\cite{WAVES} and crystals~\cite{CRYSTALS,HOLO} at 
intermediate quark chemical potentials $\mu$.  We recall that for $SU_c(2)$
the diquarks are colorless baryons and massless by the extended flavor $SU_f(4)$ symmetry~\cite{BCS}.
Most of the models lack a first principle 
description of center symmetry at finite chemical potential. This concept is usually
parametrized through a given  effective potential for the Polyakov line as in the 
Polyakov-Nambu-Jona-Lasinio models~\cite{PNJL}.  We recall that current and first principle lattice simulations
at finite chemical potential are still plagued by  the sign problem~\cite{LATTICE}, with some progress on the
bulk thermodynamics~\cite{FODOR}.

In section 2 we detail the model for two colors. 
By using a series of fermionization and bosonization techniques we show how the
3-dimensional effective action can be constructed to accommodate for the light quarks
at finite $\mu$. In section 3, we show that the equilibrium state at finite $T,\mu$ 
supports center symmetry but competing quark-antiquark or quark-quark pairing.  
In section 4, we generalize the results to arbitrary colors $N_c$.
Our conclusions are in section 5. In Appendix  A we briefly discuss the transition
matrix in the string gauge. In Appendix B we estimate the transition matrix element
in the hedgehog gauge. In Appendix C  we give an alternative but equivalent mean-field
formulation with a more transparent diagrammatic content.

\section{ Effective action with fermions at finite $\mu$}

\subsection{General setting}

In the semi-classical approximation, the Yang-Mills partition function is assumed to be dominated by an interacting ensemble of
instanton-dyons (anti-dyons).  For inter-particle distances large compared to their sizes -- or a very dilute ensemble --
 both the classical interactions and the one-loop effects are Coulomb-like. At 
 distances of the order of  the particle  sizes the one-loop effects are
 encoded in the geometry of the moduli space of the ensemble. For multi-dyons  a plausible moduli space was argued starting
from the KvBLL caloron~\cite{KVLL} that has a number of pertinent symmetries, among which permutation symmetry, overall charge neutrality, and clustering to KvBLL.

Specifically and for a fixed holonomy $A_4(\infty)/2\omega_0=\nu \tau^3/2$ with $\omega_0=\pi T$ and $\tau^3/2$
being the only diagonal color algebra generator, the
SU(2) KvBLL instanton (anti-instanton) is composed of a pair of dyons labeled by L, M (anti-dyons by $\overline {\rm L},\overline {\rm M}$)
in the notations of~\cite{DP}.  Generically there are $N_c-1$ M-dyons and only one twisted  L-dyon type. 
The SU(2) grand-partition function  is

\bea
{\cal Z}_{1}[T]&&\equiv \sum_{[K]}\prod_{i_L=1}^{K_L} \prod_{i_M=1}^{K_M} \prod_{i_{\bar L}=1}^{K_{\bar L}} \prod_{i_{\bar M}=1}^{K_{\bar M}}\nonumber\\
&&\times \int\,\frac{f_Ld^3x_{Li_L}}{K_L!}\frac{f_Md^3x_{Mi_M}}{K_M!}
\frac{f_Ld^3y_{{\bar L}i_{\bar L}}}{K_{\bar L}!}\frac{f_Md^3y_{{\bar M}i_{\bar M}}}{K_{\bar M}!}\nonumber\\
&&\times {\rm det}(G[x])\,{\rm det}(G[y])\,\left|{\rm det}\,\tilde{\bf T}(x,y)\right|\,\,e^{-V_{D\overline D}(x-y)}\nonumber\\
\label{SU2}
\eea
Here $x_{mi}$ and $y_{nj}$ are the 3-dimensional coordinate of the i-dyon of  m-kind
and j-anti-dyon of n-kind. Here
$G[x]$ a $(K_L+K_M)^2$ matrix and $G[y]$ a $(K_{\bar L}+K_{\bar M})^2$ matrix whose explicit form are given in~\cite{DP,DPX}.
$V_{D\bar D}$ is the streamline interaction between ${\rm D=L,M}$ dyons and ${\rm \bar D=\bar L, \bar M}$ antidyons as numerically discussed in~\cite{LARSEN}. For the SU(2) case  it is Coulombic asymptotically with a core at short distances~\cite{LIU1}.

The fermionic  ${\rm det}\,\tilde{\bf T}(x,y)$ determinant at finite chemical potential will be detailed below. 
The fugacities $f_{i}$ are related to the overall dyon density.
The dyon density $n_D$ could be extracted from lattice measurements of the caloron plus anti-caloron
densities at finite temperature in unquenched lattice simulations~\cite{CALO-LATTICE}. No  such extractions 
are currently available at finite density. In many ways, the partition function
for the dyon-anti-dyon ensemble resembles the partition function for the instanton-anti-instanton
ensemble~\cite{ALL}.

\subsection{Quark zero modes at finite $\mu$}

At finite $\mu$ the exact zero modes for the L-dyon (right) and $\overline{L}$-anti-dyon (left)  in the hedgehog gauge are defined as
$\varphi^A_\alpha=\eta^A_{\beta}\epsilon_{\beta\alpha}$ with indices $A$ for color and $\alpha$ for spinors.  The normalizable 
M-dyon zero mode are periodic at finite $\mu$. The L-dyon zero modes are anti-periodic at finite $\mu$. At finite $T,\mu$ they
play a dominant role in the instanton-dyon model with light quarks.  Keeping  in the time-dependence only
the lowest Matsubara frequencies  $\pm \omega_0$, their explicit form is

\begin{eqnarray}
\label{LZx}
&&\eta_R=\frac 12 \sum_{\xi=\pm }\alpha_\xi(r)\,{\bf S}_+(1-\xi\sigma\cdot \hat r)e^{i\xi(\omega_0x_4+\alpha_R)}\nonumber\\
&&\eta_L=\frac 12 \sum_{\xi=\pm }\alpha_\xi(r)\,{\bf S}_-(1+\xi\sigma\cdot \hat r)e^{i\xi(\omega_0x_4+\alpha_L)}
\end{eqnarray}
with

\begin{eqnarray}
\label{LZxx}
\alpha_\pm(r)=&&\frac{{\bf C}\,e^{\pm i\mu r}}{\sqrt{(v_l\omega_0r)\,{\rm sh}(v_l\omega_0 r)}}\nonumber\\
&&\left(\mp \frac{2i\mu}{v_l\omega_0} +
\left(e^{\mp 2i\mu r}-\frac{2}{e^{v_l\omega_0r}+1}\right)\right)
\end{eqnarray}
${\bf C}$  is an overall  normalization constant  and the SU(2) gauge rotation ${\bf S}_\pm$ 
satisfies

\be
{\bf S}_{\pm}({\sigma}\cdot {\hat r}){\bf S}_{\pm}^{\dagger}=\pm \sigma_3
\ee
translating from the hedgehog to the string gauge.  In (\ref{LZx}), $\alpha_{L,R}$ correspond to the 
rigid U(1) gauge rotations that leave the dyon coset invariant. We have kept them as they do not drop
in the hopping matrix elements below. The oscillating factors $e^{\pm 2i\mu r}$ are Friedel type oscillations.
For $\mu=0$, we recover the zero modes in~\cite{SHURYAK1,LIU2}. We have checked that the 
periodic M-dyon zero modes are in agreement with those obtained in~\cite{TIN2}.  
The restriction to the lowest Matsubara frequencies makes the mean-field analysis to follow reliable in the range
 $\mu/3\omega_0<1$.  Note that this truncation prevents the emergence of a Fermi-Dirac distribution.

\subsection{Fermionic determinant at finite $\mu$}

The fermionic determinant 
can be viewed as a sum of closed fermionic loops connecting all dyons and antidyons. Each link 
-- or hopping --
 between L-dyons and ${\rm \bar{L}}$-anti-dyons is described by the elements of the ``hopping
chiral matrix" $\tilde{\bf T}$

\begin{eqnarray}
\label{T12}
\tilde {\bf T}(x,y)\equiv \left(\begin{array}{cc}
0&i{\bf T}_{ij}\\
i{\bf T}_{ji}&0
\end{array}\right)
\end{eqnarray}
with dimensionality $(K_L+K_{\bar L})^2$.
Each of the entries in ${\bf T}_{ij}$ is a  ``hopping amplitude" for a fermion between
an L-dyon and an $\bar{\rm L}$-anti-dyon,  defined via the zero mode $\varphi_D$ of the dyon and the zero mode
$\varphi_{\bar D}$ (of opposite chirality) of the anti-dyon

\be
{\bf T}_{LR}=\int d^4x \varphi_{L}^{\dagger}(x)i(\partial_{4}-\mu-i\sigma\cdot\nabla )\varphi_R(x)\nonumber\\
{\bf T}_{RL}=\int d^4x \varphi_{R}^{\dagger}(x)i(\partial_{4}-\mu+i\sigma\cdot\nabla )\varphi_L(x)
\ee
And similarly for the other components.  These matrix elements can be made explicit in the hedgehog
gauge, 

\begin{eqnarray}
&&{\bf T}_{LR}=e^{i(\alpha_L-\alpha_R)}\,{\bf T}(p)-e^{-i(\alpha_L-\alpha_R)}\,{\bf T}^{*}(p)\nonumber\\
&&{\bf T}_{RL}=e^{i(\alpha_R-\alpha_L)}\,{\bf T}(p)-e^{-i(\alpha_R-\alpha_L)}\,{\bf T}^{*}(p)\nonumber\\
\end{eqnarray}
with a complex ${\bf T}(p)$ at finite $\mu$, 

\be
\label{ZEROFOUR}
{\bf T}(p) =-\frac 12\, ({\omega_0+i\mu})\,
\left(|f_1|^2-|f^\prime_2|^2\right)-{\rm Re}\left( f_1f_2^{\prime\,*}\right)
\ee
Here $f_{1,2}\equiv f_{1,2}(p)$ are the 3-dimensional Fourier transforms of $f_1(r)=\alpha_-(r)$ and $f_2(r)=\alpha_- (r)/r$.
The transition matrix elements in the string gauge are more involved. Their  explicit form is discussed in Appendix A.
Throughout, we will make use of the hopping matrix elements in the hedgehog gauge as the numerical difference 
between the two is small~\cite{LIU2} on average as we show in Appendix A.

\subsection{Bosonic fields}

Following~\cite{DP,LIU1,LIU2} the moduli
determinants in (\ref{SU2}) can be fermionized using 4 pairs of ghost fields $\chi^\dagger_{L,M},\chi_{L,M}$ for the dyons
and 4 pairs of ghost fields $\chi^\dagger_{{\bar L},{\bar M}},\chi_{{\bar L},{\bar M}}$ for the anti-dyons. The ensuing Coulomb factors from the determinants are then bosonized using 4 boson fields $v_{L,M},w_{L,M}$ for the dyons and similarly for
the anti-dyons.  The result is 

\bea
&&S_{1F}[\chi,v,w]=-\frac {T}{4\pi}\int d^3x\nonumber\\
&&\left(|\nabla\chi_L|^2+|\nabla\chi_M|^2+\nabla v_L\cdot \nabla w_L+\nabla v_M\cdot \nabla w_M\right)+\nonumber\\
&&\left(|\nabla\chi_{\bar L}|^2+|\nabla\chi_{\bar M}|^2+\nabla v_{\bar L}\cdot \nabla w_{\bar L}+\nabla v_{\bar M}\cdot \nabla w_{\bar M}\right)
\label{FREE1}
\eea
For the interaction part $V_{D\bar D}$, we note that
the pair Coulomb interaction in (\ref{SU2}) between the dyons and anti-dyons can also be bosonized using
standard methods~\cite{POLYAKOV,KACIR}  in terms of $\sigma$ and $b$ fields.   As a result each dyon species acquire additional
fugacity factors such that

\be
M:e^{-b-i\sigma}\,\,\,\,\, L:e^{b+i\sigma}\,\,\,\,\, \bar M: e^{-b+i\sigma}\,\,\,\,\, \bar L :e^{b-i\sigma}
 \ee
Therefore, there is an additional contribution to the free part (\ref{FREE1})

\be
S_{2F}[\sigma, b]=\frac T{8} \int d^3x\, \left(\nabla b\cdot\nabla b+ \nabla\sigma\cdot\nabla\sigma\right)
\label{FREE2}
\ee
and the interaction part is now

\bea
&&S_I[v,w,b,\sigma,\chi]=-\int d^3x \nonumber\\
&&e^{-b+i\sigma}f_M\left(4\pi v_m+|\chi_M    -\chi_L|^2+v_M-v_L\right)e^{w_M-w_L}+\nonumber\\
&&e^{+b-i\sigma}f_L\left(4\pi v_l+|\chi_L    -\chi_M|^2+v_L-v_M\right)e^{w_L-w_M}+\nonumber\\
&&e^{-b-i\sigma}f_{\bar M}\left(4\pi v_{\bar m}+|\chi_{\bar M}    -\chi_{\bar L}|^2+v_{\bar M}-v_{\bar L}\right)e^{w_{\bar M}-w_{\bar L}}+\nonumber\\
&&e^{+b+i\sigma}f_{\bar L}\left(4\pi v_{\bar l}+|\chi_{\bar L}    -\chi_{\bar M}|^2+v_{\bar L}-v_{\bar M}\right)e^{w_{\bar L}-w_{\bar M}}
\label{FREE3}
\eea
without the fermions. We now show the minimal modifications to (\ref{FREE3}) when the fermionic determinantal
interaction is present.

\subsection{Fermionic fields}

 To fermionize the determinant
and for simplicity, consider first the case of 1 flavor an 1 Matsubara frequency, and 
define the additional Grassmanians $\chi=(\chi^i_1,\chi^j_2)^T$ with $i,j=1,.., K_{L,\bar L}$ and

\be
\left|{\rm det}\,\tilde{\bf T}\right| =\int   D[\chi]\,\, e^{\,\chi^\dagger \tilde {\bf T} \, \chi}
\label{TDET}
\ee
We can re-arrange the exponent in (\ref{TDET}) by defining  a Grassmanian source $J(x)=(J_1(x),J_2(x))^T$ with

\be
J_1(x)=\sum^{K_L}_{i=1}\chi^i_1\delta^3(x-x_{Li})\nonumber\\
J_2(x)=\sum^{K_{\bar L}}_{j=1}\chi^j_2\delta^3(x-y_{\bar L j})
\label{JJ}
\ee
and by introducing 2 additional fermionic fields  $ \psi(x)=(\psi_1(x),\psi_2(x))^T$. Thus

\be
e^{\,\chi^\dagger \tilde {\bf T}\,\chi}=\frac{\int D[\psi]\,{\rm exp}\,(-\int\psi^\dagger \tilde {\bf G}\, \psi +
\int J^\dagger \psi + \int\psi^\dagger J)}{\int d
D[\psi]\, {\rm exp}\,(-\int \psi^\dagger \tilde {\bf G} \,\psi) }
\label{REFERMIONIZE}
\ee
with $\tilde{\bf G}$ a $2\times 2$ chiral block matrix

\begin{eqnarray}
 \tilde {\bf G}= \left(\begin{array}{cc}
0&-i{\bf G}(x,y)\\
-i{\bf G}(x,y)&0
\end{array}\right)
\label{GG}
\end{eqnarray}
with entries ${\bf TG}={\bf 1}$. The Grassmanian source contributions in (\ref{REFERMIONIZE}) generates a string
of independent exponents for the L-dyons and $\bar{\rm L}$-anti-dyons

\begin{eqnarray}
\prod^{K_L}_{i=1}e^{\chi_1^i\dagger \psi_{1}(x_{Li})+\psi_1^\dagger(x_{Li})\chi_1^i}\nonumber \\ \times
\prod^{K_{\bar L}}_{j=1}e^{\chi_2^j\dagger \psi_{2}(y_{\bar L j})+\psi_2^\dagger(y_{\bar L j})\chi_2^j}
\label{FACTOR}
\end{eqnarray}
The Grassmanian integration over the $\chi_i$ in each factor in (\ref{FACTOR}) is now readily done to yield

\be
\prod_{i}[-\psi_1^\dagger\psi_1(x_{Li})]\prod_j[-\psi_2^\dagger\psi_2(y_{\bar L j})]
\label{PLPR}
\ee
for the L-dyons and $\bar {\rm L}$-anti-dyons.
The net effect of the additional fermionic determinant in (\ref{SU2}) is to shift the L-dyon
and $\bar{\rm L}$-anti-dyon fugacities in (\ref{FREE3}) through

\bea
f_L\rightarrow -f_L\psi_1^\dagger\psi_1\equiv -f_L\psi^\dagger\gamma_+\psi\nonumber\\
f_{\bar L}\rightarrow -f_{\bar L}\psi_2^\dagger\psi_2\equiv -f_{\bar L}\psi^\dagger\gamma_-\psi
\label{SUB}
\eea
where we have now identified the chiralities through $\gamma_\pm=(1\pm \gamma_5)/2$.
The fugacities $f_{M,\bar M}$ are left unchanged since they do not develop zero modes.

The result (\ref{SUB}) generalizes to arbitrary number of flavors $N_f$ and two Matsubara frequencies 
labeled by $i, j=\pm$ through the substitution

\bea
f_L\rightarrow 
\prod_{f=1}^{N_f}\prod_{ i,j=\pm} \psi^\dagger_f(i_f)\gamma_+\psi_f (j_f)\,\delta\left(\sum_f(i_f+j_f)\right)\nonumber\\
f_{\bar L}\rightarrow 
\prod_{f=1}^{N_f} \prod_{ i,j=\pm} \psi^\dagger_f(i_f)\gamma_-\psi_f (j_f)\,\delta\left(\sum_f(i_f+j_f)\right)
\label{SUBf}
\eea

\subsection{Resolving the constraints}

In terms of (\ref{FREE1}-\ref{FREE3})  and the substitution
(\ref{SUB}), the dyon-anti-dyon partition function (\ref{SU2})
for finite $N_f$ can be exactly re-written as an interacting
effective field theory in 3-dimensions,

\bea
{\cal Z}_{1}[T]\equiv &&\int D[\psi]\,D[\chi]\,D[v]\,D[w]\,D[\sigma]\,D[b]\,\nonumber\\&&\times
e^{-S_{1F}-S_{2F}-S_{I}-S_\psi}
\label{ZDDEFF}
\eea
with the additional $N_f=1$ chiral fermionic contribution $S_\psi=\psi^\dagger\tilde{\bf G}\,\psi$.
Since the effective action in (\ref{ZDDEFF}) is linear in the $v_{M,L,\bar M,\bar L}$, the latters
integrate to give the following constraints

\bea
&&-\frac{T}{4\pi}\nabla^2w_M+f_M  e^{w_M-w_L}\nonumber\\&&
-f_L\prod_f \psi_f^\dagger\gamma_+\psi_f\,\e^{w_L-w_M}=\frac {T}{4\pi}\nabla^2(b-i\sigma)\nonumber\\
&&-\frac{T}{4\pi}\nabla^2w_L-f_Me^{w_M-w_L}
\nonumber\\&&
+f_L\prod_f \psi_f^\dagger\gamma_+\psi_f\,\e^{w_L-w_M}=0
\label{DELTA}
\eea
and similarly for the anti-dyons with $M,L, \gamma_+ \rightarrow \overline M, \overline L, \gamma_-$.
To proceed further the formal classical solutions to the constraint equations or $w_{M,L}[\sigma, b]$
should be inserted back into the 3-dimensional effective action. The result is

\bea
{\cal Z}_{1}[T]=\int D[\psi]\,D[\sigma]\,D[b]\,e^{-S}
\label{ZDDEFF1}
\eea
with the 3-dimensional effective action

\bea
S=&&S_F[\sigma, b]+\int d^3x\,\sum_f \psi_f^\dagger \tilde{\bf G} \psi_f\\
&&-4\pi f_M v_m\int d^3x\,( e^{w_M-w_L}+e^{w_{\bar M}-w_{\bar L}} ) \nonumber \\
&& +4\pi f_Lv_l\int d^3x\,\prod_f \psi_f^\dagger\gamma_+\psi_f\,e^{w_L-w_M}\nonumber\\
&&+4\pi f_{\bar L}v_l\int d^3x \,\prod_f \psi_f^\dagger \gamma_-\psi_f\,e^{w_{\bar L}-w_{\bar M}}\nonumber
\label{NEWS}
\eea
Here $S_F$ is $S_{2F}$ in (\ref{FREE2}) plus additional contributions resulting from the $w_{M,L}(\sigma, b)$ solutions
to the constraint equations (\ref{DELTA}) after their insertion back.  This procedure for the linearized approximation of the constraint
was discussed in~\cite{LIU1,LIU2} for the case without fermions.

\section{Equilibrium state}

To analyze the ground state and the fermionic fluctuations we  bosonize the fermions
in (\ref{ZDDEFF1})  by introducing the identities

\bea
\label{deltax}
&&\int D[\Sigma_1]\,\delta\left(\psi^\dagger_f(x)\psi_f(x)+4\Sigma_1(x)\right)={\bf 1}\\
&&\int D[\Sigma_2]\,\delta\left(
\frac 12\left(\epsilon_{fg}\psi^T_f(x)\psi_g(x)-{\rm c.c.}\right)+4i\Sigma_2(x)\right)={\bf 1}\nonumber
\eea
and re-exponentiating them to obtain

\bea
{\cal Z}_{1}[T]=\int D[\psi]\,D[\sigma]\,D[b]\,D[\vec\Sigma]\,D[\vec\lambda]\,
e^{-S-S_C}\nonumber\\
\label{ZDDEFF2}
\eea
with

\bea
\label{SCx}
&&-S_C=\int d^3x \,i\lambda_1(x)(\psi_f^{\dagger}(x)\psi_f(x)+4\Sigma_1(x))\\ 
&&+\int d^3x \,{i\lambda_2}(x)\left(
\frac 12\left(\epsilon_{fg}\psi^T_f(x)\psi_g(x)-{\rm c.c.}\right)+4i\Sigma_2(x)\right)\nonumber
\eea
The ground state is parity even so that $f_{L,M}=f_{\bar L, \bar M}$.
By translational invariance, the SU(2) ground state corresponds to constant $\sigma, b, \vec\Sigma, \vec\lambda$.
We will seek the extrema of (\ref{ZDDEFF2}) with finite condensates in the mean-field approcimation, i.e.

\bea
\label{DEFCC}
&&\left<\psi^\dagger_f(x)\psi_g(x)\right>=-2\delta_{fg}\Sigma_1\nonumber\\
&&\left<\psi^T_f(x)\psi_g(x)\right>=-2i\epsilon_{fg}\Sigma_2
\eea
With this in mind, the classical solutions to the constraint equations (\ref{DELTA}) are also constant

\bea
\label{SU2SOL}
f_M  e^{w_M-w_L}
=f_L\left<\prod_f \psi_f^\dagger\gamma_+\psi_f\right>\,\e^{w_L-w_M}
\eea
with

\bea
\label{SU2SOLx}
\left<\prod_f \psi_f^\dagger\gamma_+\psi_f\right>=\left(\Sigma_1^2+\Sigma_2^2\right)\equiv {\vec\Sigma}^2
\eea
and similarly for the anti-dyons. The expectation values in (\ref{SU2SOL}-\ref{SU2SOLx}) 
are carried in  (\ref{ZDDEFF2}) in the mean-field approximation through Wick contractions. 
Here we note that both the chiral  pairing ($\Sigma_1$) and diquark pairing ($\Sigma_2$) 
are of equal strength in the instanton-dyon liquid model. The chief reason is that the pairing
mechanism goes solely through the KK- or L-zero modes which are restricted to the affine
root of the color group. With this in mind,  the solution to (\ref{SU2SOL}) is

\be
e^{w_M-w_L}=|\vec\Sigma|\,\left(\frac{f_L}{f_M}\right)^{\frac 12}
\ee
and similarly for the anti-dyons.

\subsection{Effective potential}

The effective potential ${\cal V}$ for constant fields follows from (\ref{ZDDEFF2})
by enforcing the delta-function constraint (\ref{NEWS}) before variation (strong constraint) 
and parity

\bea
-{\cal V}/\mathbb{V}_3=&&-4\,\vec\lambda\cdot\vec\Sigma\nonumber\\&&+4\pi f_M v_m\,( e^{w_M-w_L}+e^{w_{\bar M}-w_{\bar L}} ) \nonumber \\
&& +4\pi f_Lv_l\,{\vec\Sigma}^2\,(e^{w_L-w_M}+e^{w_{\bar L}-w_{\bar M}})
\label{POT}
\eea
after shifting $\lambda_1\rightarrow i\lambda_1$ for convenience, with $\mathbb{V}_3$ the 3-volume.
For fixed holonomies $v_{m,l}$, the constant $w^\prime$s are real by (\ref{DELTA})
as all right hand sides vanish, and the extrema of (\ref{POT}) occur for

\bea
e^{w_M-w_L}=\pm |\vec\Sigma |\,\sqrt{f_Lv_l/f_Mv_m}\nonumber\\
e^{w_{\bar M}-w_{\bar L}}=\pm |\vec\Sigma |\,\sqrt{ f_{L}v_{\bar l}/f_Mv_{\bar m}}
\label{EXT}
\eea
(\ref{EXT}) are  consistent with  (\ref{SU2SOL}) only if
$v_l=v_m=1/2$ and $v_{\bar l}=v_{\bar m}=-1/2$. That is for confining holonomies or a center
symmetric ground state. Thus

\be
- {\cal V}/\mathbb{V}_3=\alpha\,|\vec\Sigma|-4\,\vec\lambda\cdot\vec\Sigma 
\label{SU2POT}
\ee
with $\alpha=4\pi\sqrt{f_Lf_M}$.
We note that for $\vec\Sigma=\vec0$ there are no solutions to the extrema equations. 
Since $\vec\Sigma=\vec0$ means a zero chiral or quark condensate
(see below), we conclude that in this model of the  dyon-anti-dyon liquid with light quarks, 
center symmetry is restored only if both the chiral and superconducting condensates vanish.

\subsection{Gap equations}

For the vacuum solution, the auxiliary field $\vec\lambda$ is also a constant.
The fermionic fields in (\ref{ZDDEFF2}) can be integrated out. The result is
a new contribution to the potential (\ref{SU2POT})

\bea
&&- {\cal V}/\mathbb{V}_3\rightarrow\alpha\,|\vec\Sigma|-4\vec\lambda\cdot\vec\Sigma \nonumber\\
&&+2\int \frac{d^3p}{(2\pi)^3}{\rm ln}\,
\left(\left(1+{\vec\lambda}^2|{\bf T}(p)|^2\right)^2-4\lambda_1^2|{\rm Im}{\bf T}(p)|^2\right)\nonumber\\
\label{SU2POT1}
\eea

The saddle point  of (\ref{SU2POT1}) in $\vec\Sigma$  is achieved for  parallel vectors

\be
\vec\lambda=\frac \alpha4 \frac{\vec\Sigma}{|\vec \Sigma|}\equiv \lambda\, \left({\rm cos}\,\theta, {\rm sin}\,\theta\right)
\label{L1}
\ee
Inserting (\ref{L1}) into the effective potential (\ref{SU2POT1}) yields

\bea
\label{E1}
&&- {\cal V}/\mathbb{V}_3=2\int \frac{d^3p}{(2\pi)^3}\\
&&\times{\rm ln}\,
\left(\left(1+\lambda^2|{\bf T}(p)|^2\right)^2-4\lambda^2{\rm cos}^2\theta\,|{\rm Im}{\bf T}(p)|^2
\right)\nonumber
\eea
with $\lambda=\alpha/4$ now  fixed.
(\ref{E1}) admits 4 pairs of discrete extrema satisfying $\delta {\cal V}/\delta\theta=0$ with  ${\rm cos}\,\theta=0,1$.
 The extrema carry the pressure per 3-volume

\bea
\label{E1x}
&&- {\cal V}_{0,1}/\mathbb{V}_3=2\int \frac{d^3p}{(2\pi)^3}\\
&&\times{\rm ln}\,
\left(\left(1+\lambda^2|{\bf T}(p)|^2\right)^2-4\lambda^2 (0,1) |{\rm Im}{\bf T}(p)|^2
\right)\nonumber
\eea

We note ${\rm Im}{\bf T}=0$ in (\ref{SU2POT1}) for $\mu=0$. The effective potential has manifest extended flavor $SU_f(4)$ symmetry
which is spontaneously broken  by the saddle point (\ref{L1}).  Since zero $\mu$ cannot support the breaking of 
$U(1)_V$, this phase is characterized by a finite chiral condensate and a zero diquark condensate. For 
$\mu\neq 0$, we have ${\rm Im}{\bf T}\neq 0$ in (\ref{SU2POT1}). The effective potential loses manifest $SU_f(4)$ symmetry. 
While the saddle point (\ref{L1}) indicates the possibility of either a chiral or diquark condensate, (\ref{E1x}) shows
that the diquark phase is favored by a larger pressure since ${\cal V}_0>{\cal V}_1$. The $\mu>0$ is a superconducting phase 
of confined baryons.

The chiral and diquark condensates follow from the definitions (\ref{DEFCC})  and the saddle point
(\ref{L1}), which are

\be
\left(\frac{\left<\bar q q\right>}T, -\frac{\left<qq\right>}T\right)=-2(\lambda_1, \lambda_2)
\label{CC0}
\ee
For $\mu=0$ we have $\lambda_2=0$ and $\left<\bar q q\right>/T=-\alpha/2$, while for $\mu\neq 0$ we have $\lambda_1=0$
and ${\left<qq\right>}/T=\alpha/2$, with $\alpha=4\pi\sqrt{f_Lf_M}$  which is independent of $\mu$.

\subsection{Constituent quark mass and scalar gap}

\begin{figure}[h!]
  \begin{center}
  \includegraphics[width=7cm]{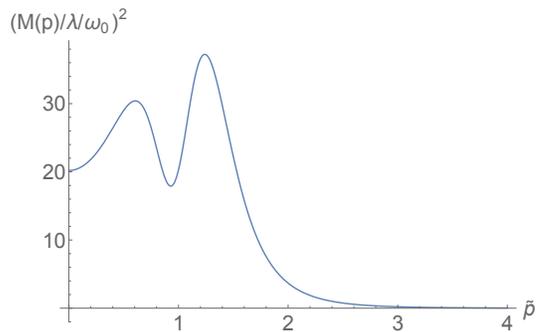}
   \caption{The squared momentum dependent quark constituent mass $(\omega_0 M(p)/\lambda)^2$ versus $\tilde p=p/\omega_0$
   for $\mu/\omega_0=1$. }
     \label{fig_mass1}
  \end{center}
\end{figure}

In the paired phase with $\lambda_1=0$, the momentum-dependent constituent quark mass $M(p)$  can be defined using 
 the determinant (\ref{E1x}) to be

\be
M(p)=\lambda\,\left({\omega_0^2+p^2}\right)^{\frac 12} \,|{\bf T}(p)|
\label{MASS}
\ee
In Fig.~\ref{fig_mass1} we show the behavior of the dimensionless mass ratio $(M(p)/\lambda/\omega_0)^2$
as a function of $p/\omega_0$.  The oscillatory behavior is a remnant of the Friedel oscillation
noted earlier. We note that (\ref{MASS}) through (\ref{SCx}-\ref{DEFCC}) satisfies

\be
\int \frac{d^3p}{(2\pi)^3}\frac{M^2(p)}{\omega^2_0+p^2+M^2(p)}= \frac{n_D}8
\label{GAPX}
\ee
with $n_D=8\pi\sqrt{f_Lf_M}\Sigma$. 




The superconducting mass gap $\Delta(0)/2$ can be obtained by fluctuating along the modulus of 
the paired quark $qq$. This is achieved through a small and local scalar 
deformation  of the type $\lambda_2(x)\approx \lambda (1+is(x))$, for which the effective action to quadratic
order is

\be
{\bf S}[s]=\frac{2N_f}{2f_s}\int\frac{d^3p}{(2\pi)^3}s(p)\,\frac 1{\Delta_s(p)}\,s(-p)
\ee
For $x\neq 1$, the scalar propagator is ($p_{\pm}=q\pm\frac{p}{2}$)

\be
\frac 1{\Delta_s(p)}=\frac{x}{1-x}\frac{n_D}{N_c}+\int \frac{d^3 q}{(2\pi)^3}\frac{(p_-M_++p_+M_-)^2}{(p_+^2+M_+^2)(p_-^2+M_-^2)}\\
\ee
while for $x=1$ it is

\be
\frac 1{\Delta_s(p)}=2\int\frac{d^3 q}{(2\pi)^3}\frac{M_+M_-(p_+p_--M_-M_+)}{(p_+^2+M_+^2)(p_-^2+M_-^2)}
\ee
Here we have defined $M_\pm \equiv \lambda |{\bf T}(p_\pm)|$, and therefore 
$\Delta(0)=\Delta_s(0)/\Delta_s^\prime(0)$.

\section{Generalization to $SU_c(N_c)\times SU_f(N_f)$}

For general $N_c$ with $x=N_f/N_c$,  the pairing in (\ref{DEFCC}) involves only those color indices 
commensurate with the affine root of $SU_c(N_c)$ through their corresponding KK- or 
L-zero modes. This leaves $(N_c-2)$ colors with energy $\omega_0$ unpaired. As a result,
the non-perurbative pressure per unit 3-volume of the paired colored quarks (\ref{SU2POT1}) is now changed to

\bea
&&- {\cal V}/\mathbb{V}_3=\alpha\,|\vec\Sigma|^x-4\vec\lambda\cdot\vec\Sigma \nonumber\\
&&+N_f\int \frac{d^3p}{(2\pi)^3}{\rm ln}\,
\left(\left(1+{\vec\lambda}^2|{\bf T}(p)|^2\right)^2-4\lambda_1^2|{\rm Im}{\bf T}(p)|^2\right)\nonumber\\
\label{SU2POTNC}
\eea
with now instead $\alpha=4\pi(f_Lf_M^{N_c-1})^{\frac 1{N_c}}$.  The extrema in $\vec \Sigma$ still
yield parallel vectors

\bea
&&\vec\lambda =\lambda ({\rm cos }\theta, {\rm sin}\theta) \nonumber\\
&&\vec\Sigma =\Sigma ({\rm cos}\theta, {\rm sin}\theta) 
\eea
for which (\ref{SU2POTNC}) simplifies

\bea
&&- {\cal V}/\mathbb{V}_3=\alpha\,\Sigma^x-4\lambda\Sigma \nonumber\\
&&+N_f\int \frac{d^3p}{(2\pi)^3}{\rm ln}\,
\left(\left(1+{\lambda}^2|{\bf T}(p)|^2\right)^2-4\lambda^2\,{\rm cos}^2\theta\,|{\rm Im}{\bf T}(p)|^2\right)\nonumber\\
\label{SU2POTNCx}
\eea
The saddle point in $\Sigma$ gives

\be
\lambda=\frac \alpha4 { x\Sigma^{x-1}}
\ee
while the saddle point in $\theta$ gives ${\rm cos}\,\theta=0, 1$. The latter yields the respective pressure per volume

\bea
&&- {\cal V}_{0,1}/\mathbb{V}_3=\alpha\left(\frac{4}{\alpha x}\right)^{\frac{x}{x-1}}(1-x)\lambda^{\frac{x}{x-1}}\nonumber\\
&&+N_f{\rm ln}\,
\left(\left(1+\lambda^2|{\bf T}(p)|^2\right)^2-4\lambda^2(0,1)|{\rm Im}{\bf T}(p)|^2
\right)\nonumber
\label{SUPOTNCxx}
\eea
For $\mu=0$, we have ${\cal V}_0={\cal V}_1$ and both the chiral and diquark phase are 
degenerate. Since the $\mu=0$ phase cannot break $U(1)_V$, the chiral phase with a pion
as a Golstone mode is favored. For $\mu>0$, ${\cal V}_0>{\cal V}_1$, the diquark phase
is favored by the largest pressure. Since the phase is center symmetric, this 
implies that the  baryon chemical potential $\mu_B$
satisfies $\mu_B\equiv N_c\mu>(N_c-2)\omega_0$. The transition from the
chiral phase to the diquark phase is first order.

\section{Conclusions}

We have extended the mean field treatment of the SU(2)  instanton-dyon model 
with light quarks in~\cite{LIU1} to finite chemical potential $\mu$. In Euclidean space, finite
$\mu$ enters through $i\mu$ in the Dirac equation. The anti-periodic KK- or  L-dyon
zero modes are calculated for the lowest Matsubara frequencies. The delocalization 
occur only through the KK- or L-dyon zero modes  which implies that the diquark
pairing and the chiral pairing have equal strength whatever $N_c$. Therefore,
the instanton-dyon liquid may not support chiral density waves~\cite{WAVES}. 
The di-quark phase  is  favored  for $\mu>(1-2/N_c)\omega_0$ under the additional stricture
of center symmetry.
A useful improvement on this work would be a re-analysis of the KK- or L-zero modes 
including all Matsubara frequencies.

\section{Acknowledgements}

This work was supported by the U.S. Department of Energy under Contract No.
DE-FG-88ER40388.

\section{Appendix A: Fermionic hopping in the string gauge at finite $\mu$}

In this Appendix we detail the form of the hopping matrix in the string gauge.
We will show that the difference with the  hopping matrix element in the
hedgehog gauge (\ref{ZEROFOUR})  used in the main text is (numerically) small. 

We transform the L-zero modes in hedgehog gauge (\ref{LZx}) to the string gauge 
using the $(\theta, \phi)$ polar parametrization of ${\bf S}_\pm$,

\bea
\label{AZZ1}
\psi^{a=1}_{\rm L}=&&e^{-i\omega_0x_4}\left(-{\rm sin}\frac{\theta}{2}e^{-i\phi},+{\rm cos}\frac{\theta}{2}\right)\alpha_+(r)\nonumber\\
\psi^{a=2}_{\rm L}=&&e^{+i\omega_0x_4}\left(-{\rm cos}\frac{\theta}{2},-{\rm sin}\frac{\theta}{2}e^{+i\phi}\right)\alpha_-(r)
\eea
and similarly for the $\bar {\rm L}$-dyon

\bea
\label{AZZ2}
\psi^{a=1}_{\bar{\rm  L}}=&&e^{-i\omega_0x_4}\left(-{\rm cos}\frac{\theta}{2},-{\rm sin}\frac{\theta}{2}e^{+i\phi}\right)\alpha_-(r)\nonumber\\
\psi^{a=2}_{\bar {\rm L}}=&&e^{+i\omega_0x_4}\left(-{\rm sin}\frac{\theta}{2}e^{-i\phi},+{\rm cos}\frac{\theta}{2}\right)\alpha_+(r)
\eea
with  $\alpha_{\pm}(r)$ defined in (\ref{LZxx}).
In terms of (\ref{AZZ1}-\ref{AZZ2}) the hopping matrix element (\ref{T12}) involves the relative 
angular orientation $\theta$ (not to be confused with $\theta$ used in the text). It is in general numerically involved.

To gain further  insights and simplify physically the numerical analysis, 
let $l_{xy}$ be the line segment connecting $x$ to $y$ in (\ref{T12}) and let $z$ lies on it.
Since the zero modes decay exponentially, the dominant z-contribution to the integral in (\ref{T12}) stems 
from those $z$ with the smallest $|x-z|+|y-z|$ contribution. Using rotational symmetry, we can set 
$x=0$ and $y=(r,\theta,0)$ in spherial cordinates. The dominant  contributions are from 
$\theta_{z-x}=\theta,\phi_{z-x}=0$, and $\theta_{z-y}=\pi-\theta,\phi_{z-y}=-\pi$ which can be viewed as constant in the integral. 
With this in mind, (\ref{T12}) in string gauge reads

\bea
\label{TIJS}
&&-{\bf T}_{LR}^+(x-y)=\nonumber\\
&&\frac{\omega_0+i\mu}{2}\int d^3z\, \alpha_+^*(|x-z|)\alpha_+(|y-z|) \nonumber\\
&&-\frac{1}{2}\left(1+\frac{{\rm cos}^2 \theta-{\rm cos}\theta}{2}\right)\nonumber\\
&&\times {\rm Re}\int d^3z\, \alpha^*_+(|x-z|)\frac{\alpha^\prime_+(|y-z|)+\alpha_+(|y-z|)}{|y-z|}\nonumber\\
\eea
In a large ensemble of dyons and anti-dyons, we have on average 
$\left<{\rm cos} \theta\right>=0$ and $\left<{\rm cos}^2\theta\right>=\frac{1}{2}$.  Thus, 

\bea
\label{TIJSx}
&&{\bf T}_{LR}^+(x-y)=\nonumber\\
&&\frac{\omega_0}{2}\int d^3z \,\alpha_+^*(|x-z|)\alpha_+(|y-z|) \nonumber\\
&&-\frac {5}{8} {\rm Re} \int d^3z\,\alpha^*_+(|x-z|)\frac{\alpha^\prime_+(|y-z|)+\alpha_+(|y-z|)}{|y-z|)}\nonumber\\
\eea
in the string gauge. Its Fourier transform is

\be
{\bf T}(p)\approx-\frac{1}{2}\left((\omega_0+ i \mu)|\alpha_+(p)|^2-\frac{5}{4}{\rm Re}(\alpha_+^{*}(p)\tilde \alpha_{+}(p))\right)\nonumber\\
\label{ATP}
\ee
with $\tilde \alpha(r)=(r\alpha_{+}(r))^\prime/r$.
(\ref{ATP})  is to be compared to (\ref{ZEROFOUR}) in the hedgehog gauge.  The dominant contribution in (\ref{ATP}) 
is due to the first contribution $|\alpha_+|^2$ which is common to both gauge fixing. A similar observation was made 
in~\cite{LIU2} for the case of  $\mu=0$.

\section{Appendix B: Estimate of the fermionic hopping in the hedgehog gauge at finite $\mu$}

In this Appendix we we will estimate the fermionic hopping matrix element  (\ref{ZEROFOUR})
by using the asymptotic form of the L-dyon zero mode at finite $\mu$ (\ref{LZx}-\ref{LZxx}). 
Throughout we  will use the dimensionless redefinitions $\mu\rightarrow \mu/\omega_0$ 
and $p\rightarrow p/\omega_0$. The normalization in  (\ref{LZx}) is fixed with 

\be
{\bf C}=\omega_0^{\frac 32}\left(8\pi(1+4\mu^2)\right)^{\frac 12}
\eea
With this in mind,  (\ref{ZEROFOUR}) reads

\bea
{\bf T}(p)=-\frac{\pi}{\omega_0^2}\frac{1}{(1+4\tilde \mu^2)}
\left((1+i\mu){\mathbb  F}_1(p)+{\mathbb F}_2(p)\right)
\eea
with

\bea
&&{\mathbb F}_1(p)=a^2_1(p)-A_1^{\prime 2}(p)+a_2^2(p)-A_2^{\prime 2}(p)\nonumber\\
&&{\mathbb F}_2 (p)=2p\left(a_1(p)A_1^\prime(p)+a_2(p)A_2^\prime(p)\right)
\eea

\bea
a_1(p)=\frac{1}{p}\int_{0}^{\infty}\sqrt{x}\,{\rm sin}(px)\left(2\mu \,{\rm sin}(\mu x)+{\rm cos}(\mu x)\right)\nonumber\\
a_2(p)=\frac{1}{p}\int_{0}^{\infty}\sqrt{x}\,{\rm sin}(px)\left(2\mu \,{\rm cos}(\mu x)-{\rm sin}(\mu x)\right)\nonumber\\
A_1(p)=\frac{1}{p}\int_{0}^{\infty}\frac 1{\sqrt{x}}\,{\rm sin}(px)\left(2\mu \,{\rm sin}(\mu x)+{\rm cos}(\mu x)\right)\nonumber\\
A_2(p)=\frac{1}{p}\int_{0}^{\infty}\frac 1{\sqrt{x}}\,{\rm sin}(px)\left(2\mu \,{\rm cos}(\mu x)-{\rm sin}(\mu x)\right)
\eea
More explicitly, define

\bea
&&a(p)=\frac{\sqrt{2 \pi } \sin \left(\frac{3}{2} \tan ^{-1}(2 p)\right)}{\left(4 p^2+1\right)^{3/4}}\nonumber\\
&&b(p)=\frac{\sqrt{2 \pi } \cos \left(\frac{3}{2} \tan ^{-1}(2 p)\right)}{\left(4 p^2+1\right)^{3/4}}\nonumber\\
&&A(p)=\frac{2 \sqrt{\pi } p}{\sqrt{4 p^2+1} \sqrt{\sqrt{4 p^2+1}+1}}\nonumber\\
&&B(p)=\frac{\sqrt{\pi } \sqrt{\sqrt{4 p^2+1}+1}}{\sqrt{4 p^2+1}}
\eea
Then we have

\bea
a_1(p)=&&\frac{1}{p}(\mu(b(p-\mu)-b(p+\mu))\nonumber\\
&&-\frac{1}{2}(a(p+\mu)+a(p-\mu))\nonumber \\
a_2(p)=&&\frac{1}{p}(\mu(a(p-\mu)+a(p+\mu))\nonumber\\
&&-\frac{1}{2}(b(p+\mu)-b(p-\mu))\nonumber \\
A_1(p)=&&\frac{1}{p}(\mu(B(p-\mu)-B(p+\mu))\nonumber\\
&&-\frac{1}{2}(A(p+\mu)+A(p-\mu))\nonumber \\
A_2(p)=&&\frac{1}{p}(\mu(A(p-\mu)+A(p+\mu))\nonumber\\
&&-\frac{1}{2}(B(p+\mu)-B(p-\mu))\nonumber \\
\eea
We note the momentum averaged hopping strengths

\bea
&&\mu=0:\qquad\int \frac{d^3p}{(2\pi)^3}|{\bf T}(p)|^2\approx \frac{4.86}{\omega_0}\nonumber\\
&&\mu=\omega_0:\qquad\int \frac{d^3p}{(2\pi)^3}|{\bf T}(p)|^2 \approx \frac{0.98}{\omega_0}
\ee
and  the typical hopping strengths at zero momentum is

\bea
&&\mu=0:\qquad|{\bf T}(0)|^2\approx \frac{307.97}{T^4}\nonumber\\
&&\mu=\omega_0:\qquad|{\bf T}(0)|^2 \approx \frac{0.20}{T^4}
\eea
We note the huge reduction in hopping at $\mu=\omega_0$.

\section{Appendix C:  Alternative effective action at finite $\mu$}

In this Appendix, we detail an alternatie mean-field analysis of the instanton-dyon
ensemble at finite $T,\mu$. The construction is more transparent for a diagrammatic
interpretation and allows for the use of many-body techniques beyond the mean-field
limit. For that, we set $N_f=2$ and define

\be
\left<\psi_f(p)\psi^{\dagger}_{g}(-p)\right>=\delta_{fg}{\bf F}_1(p)\\
\left<\psi_f(p)\psi^{T}_g(-p)\right>=i\epsilon_{fg}{\bf F}_2(p)
\ee  
with $p=(\vec p,\pm\omega_0)$ subsumed. The averaging is assumed over the
instanton-dyon ensemble, with 

\be
\Sigma_{1,2}=\frac{1}{2}{\rm Tr}\, {\bf F}_{1,2}
\ee
The Trace is carried over the dummy spin indices and momentum. 
The 3-dimensional effective action  for the momentum dependent 
spin matrices ${\bf F}_{1,2}$ in the mean-field approximation takes the 
generic form

\bea
\label{GEN}
-\Gamma[{\bf F}]=&&
\alpha\left(\left(\frac{{\rm Tr} {\bf  F}_1}{2}\right)^{2}+\left(\frac{{\rm Tr} {\bf F}_{2}}{2}\right)^{2}\right)^{\frac{1}{N_c}} \\
&&+2{\rm Tr}{\tilde{\bf G}}{\bf F}_1-{\rm Tr ln}\left({\bf  F}_2^2+{\bf F}_2{\bf F}_1^T{\bf F}_2^{-1}{\bf F}_1\right)\nonumber
\eea
The first contribution is the Hartree-Fock type contribution to the effective potential after 
minimizing with respect to $(w_M-w_L)$. The second and third contributions are
from the fermionic loop with the fermion propagator evaluated in the mean-field
approximation. We note that in the dyon ensemble both the quark-quark pairing 
and the quark-anti-quark pairing carry equal weight in the Hartree-Fock term. This
is not the case for one-gluon exchange or the instanton liquid model where the
quark-quark pairing is $1/N_c$ suppressed in comparison to the quark-anti-quark pairing.
We have checked that the saddle point equations 

\be
\frac{\delta\Gamma[{\bf F}]}{\delta {\bf F}_i(p)}=0
\ee
yield the saddle point results in the main text. 

 \vfil

\end{document}